\documentclass[aps,prl,preprint,groupedaddress,showpacs]{revtex4}

\usepackage{graphicx}
\usepackage{float}

\begin{document}

\title{Supplementary Information: Quantum phase transition in a
single-molecule quantum dot}

\author{Nicolas Roch$^1$, Serge Florens$^1$, Vincent Bouchiat$^1$, 
Wolfgang Wernsdorfer$^1$ {\&} Franck Balestro$^1$}

\affiliation{$^1$Institut N\'eel, associ\'e \`a l'UJF, CNRS, BP 166,
38042 Grenoble Cedex 9, France}

\begin{abstract}
Quantum criticality is the intriguing possibility offered by the laws of
quantum mechanics when the wave function of a many-particle physical system is
forced to evolve continuously between two distinct, competing ground states.
This phenomenon, often related to a zero-temperature magnetic phase
transition, can be observed in several strongly correlated materials such as
heavy fermion compounds or possibly high-temperature superconductors, and is
believed to govern many of their fascinating, yet still unexplained properties. 
In contrast to these bulk materials with very complex electronic structure,
artificial nanoscale devices could offer a new and simpler vista to the
comprehension of quantum phase transitions.
This long-sought possibility is demonstrated by our work in a fullerene molecular
junction, where gate voltage induces a crossing of singlet and triplet spin states
at zero magnetic field.
Electronic tunneling from metallic contacts into the $\rm{C_{60}}$ quantum 
dot provides here the necessary many-body correlations to observe a true 
quantum critical behavior.
\end{abstract}

\maketitle

\section{Experimental setup}
\label{sec:setup}

The spin $S=1/2$ Kondo effect in a $\rm{C_{60}}$ molecular junction was observed
for the first time by Yu and Natelson~\cite{Yu2004} (see also
\cite{Pasupathy2004} in the case of ferromagnetic electrodes), and
more recently by Parks {\it{et al.}}~\cite{Parks2007} using mechanically controllable break
junctions. However, to our knowledge, no electromigration 
procedure has been carried out in a dilution refrigerator with a 
high degree of filtering. The creation of nanogaps with this
technique requires minimizing the series
resistance~\cite{vanderZant2006}, which is generally incompatible
with dilution fridge wiring and filtering. To overcome this problem, we
developed a specific measurement setup described here.
We emphasise that the possibility of accessing very low
temperatures, as compared to relatively large Kondo scales, was central
to the observation of quantum critical signatures
associated to the singlet-triplet crossing in this system.

\begin{figure}
\includegraphics[width=16cm]{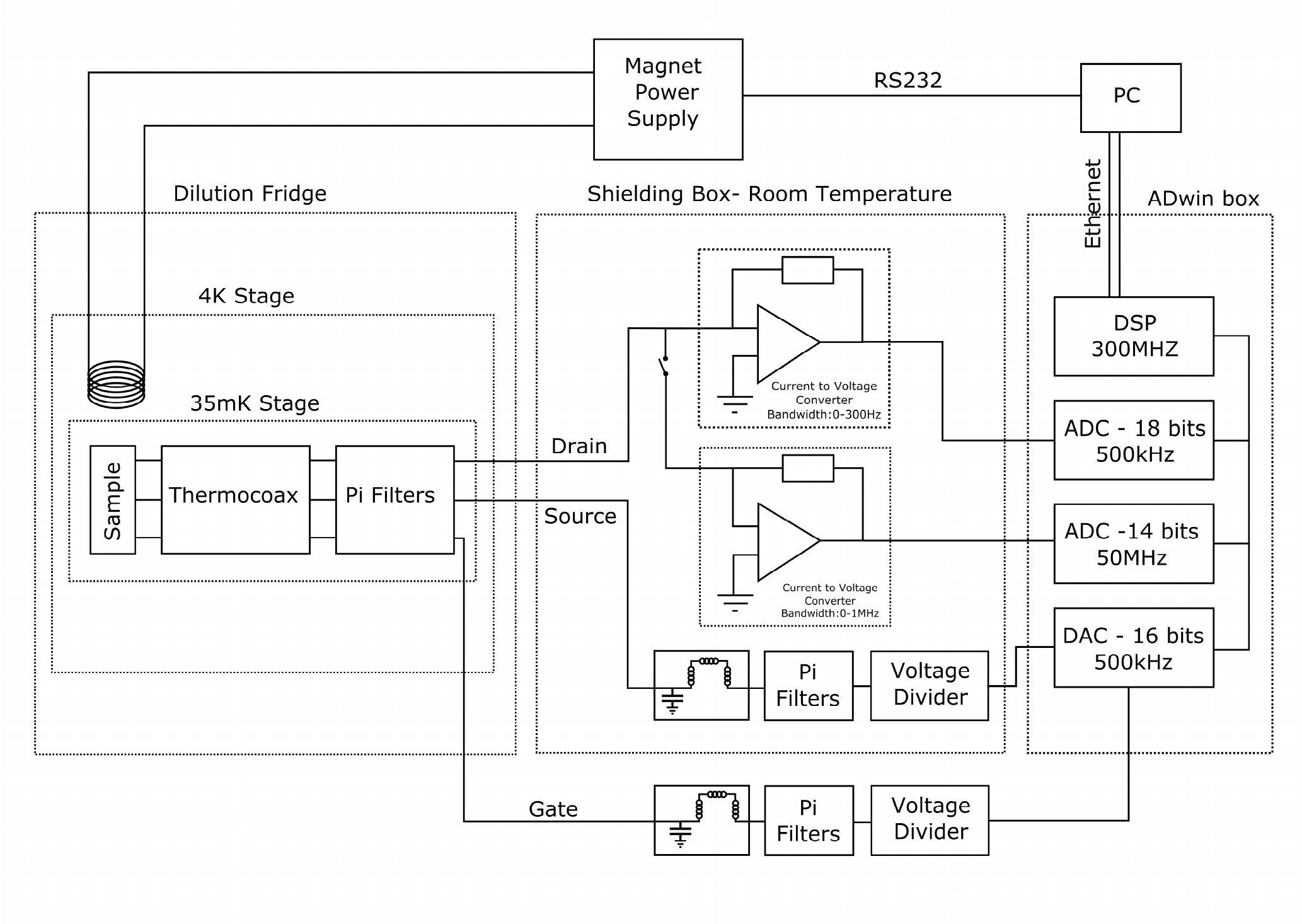}
\caption{ \label{figsetup} {\bf Simplified scheme of the experimental
setup.} See text for details.}
\end{figure}

Our experimental setup (Fig.~S.~\ref{figsetup}) is divided into two
parts. First, electromigration~\cite{Park1999} is performed at 4~K with the fast part of the
setup. As we wanted to perform such measurements in a dilution
fridge, we developed an efficient electromigration technique
since dilution wires and low-temperature filters are very resistive
and add an important series resistance to the sample (few hundreds
Ohms). Improvements of the original procedure~\cite{Park1999} have
already been reported recently~\cite{Strachan2005, Houck2005,
Esen2005, Trouwborst2006, O'Neill2007, Wu2007}. We ramp the
voltage across the junction and measure its resistance using a very
fast feedback-loop~($\rm{1.5}$~$\rm{\mu s}$) in order to set the voltage to
zero when the resistance exceeds a defined threshold, typically 
20~k$\Omega$. The fast
feed-back loop was achieved with a real-time electronics (Adwin Pro
II) and a home-built high-bandwidth current to voltage converter, as
described in Fig S.\ref{figsetup}). With this technique, we obtained
small gaps (1-2~nm) characterized by the tunnel
current measured after electromigration, without molecules, in 
previous experiments.

\def\figurename{Fig. S.}

The second step uses the low noise component of the setup to measure the
single-molecule transistor. In addition to low-temperature
filtering, we used ${\Pi}$ filters and ferrite bead filters developed at Harvard by
J. MacArthur and B. D'Urso~\cite{Marcus}. In order
to minimise
ground loops we integrated all the analogical electronics in a shielded box at
room temperature. Because of its great versatility, Adwin Pro II can
be programmed to perform DC or lock-in measurements, and apply gate
or bias voltages, thus minimizing the possibility of ground loops.
Depending on the measurements, we used an AC-excitation
between 3~$\rm{\mu V}$ and 100~$\rm{\mu V}$ for the lock-in technique.

We note that previous studies of
$\rm{C_{60}}$ quantum dots did not require the use of a dilution
refrigerator to investigate Kondo physics because the relevant energy scales
are typically an order of magnitude larger than
in carbon nanotubes or semiconducting devices, providing large Kondo temperatures of
several kelvins. However, the study of single-molecule transistors
using
low-temperature techniques (previously reserved to 2DEG systems) was certainly crucial
for unveiling the rich physics that takes place below the Kondo temperature at the
singlet-triplet transition.
Our low-temperature setup allowed a more precise investigation of
the usual spin $S=1/2$ Kondo effect in $\rm{C_{60}}$ and is presented in the next section.

\section{Fully-screened spin $S=1/2$ Kondo effect in a $\rm{C_{60}}$ quantum dot}
\label{sec:screened}

In this section we present a detailed study of the standard spin
$S=1/2$ Kondo effect observed in the Coulomb diamond associated
with an odd
excess number of electrons into the $\rm{C_{60}}$ molecule. In this particular region we
clearly observe a zero bias anomaly in the conductance, as shown
in Fig.1c of the main paper. This signature has been
widely observed in semiconducting devices~\cite{Goldhaber1998,Cronenwett1998}, carbon nanotube \cite{Nygard2000,Liang2002},
or single-molecule~\cite{Liang2002bis, Park2002,Yu2004} quantum dots. In
such strongly confined nanostructures, when the last electronic
energy level is occupied by a single electron, the quantum dot
behaves as a spin $S=1/2$ magnetic impurity. In this case the
conduction electrons in the leads are coupled antiferromagnetically
to the magnetic impurity via second order tunneling processes. When
the tunnel barriers between the dot and the electrodes are
transparent enough, so that resonant Kondo scattering can occur at low
temperature, quantum coherent transport establishes and allows the
current to flow through the dot, thus beating the Coulomb
blockade~\cite{Glazman1988, Ng1988}. When conduction
electrons form a Kondo cloud around the dot to screen its magnetic
moment, a sharp peak is created in the density of state at the
Fermi level, giving rise to a narrow resonance in the differential
conductance, which does not disperse with varying the gate voltage.
Universality is a fundamental property of the Kondo effect and a single energy scale,
associated with the Kondo temperature $T_{\rm K}$, fully describes the
physical properties at low energy. When the typical energy of a
perturbation, such as temperature, bias voltage, or magnetic field,
is higher than $T_{\rm K}$, the coherence of the system is
suppressed and the Kondo effect disappears.

\def\figurename{Fig. S.}
\begin{figure}
\includegraphics[width=16cm]{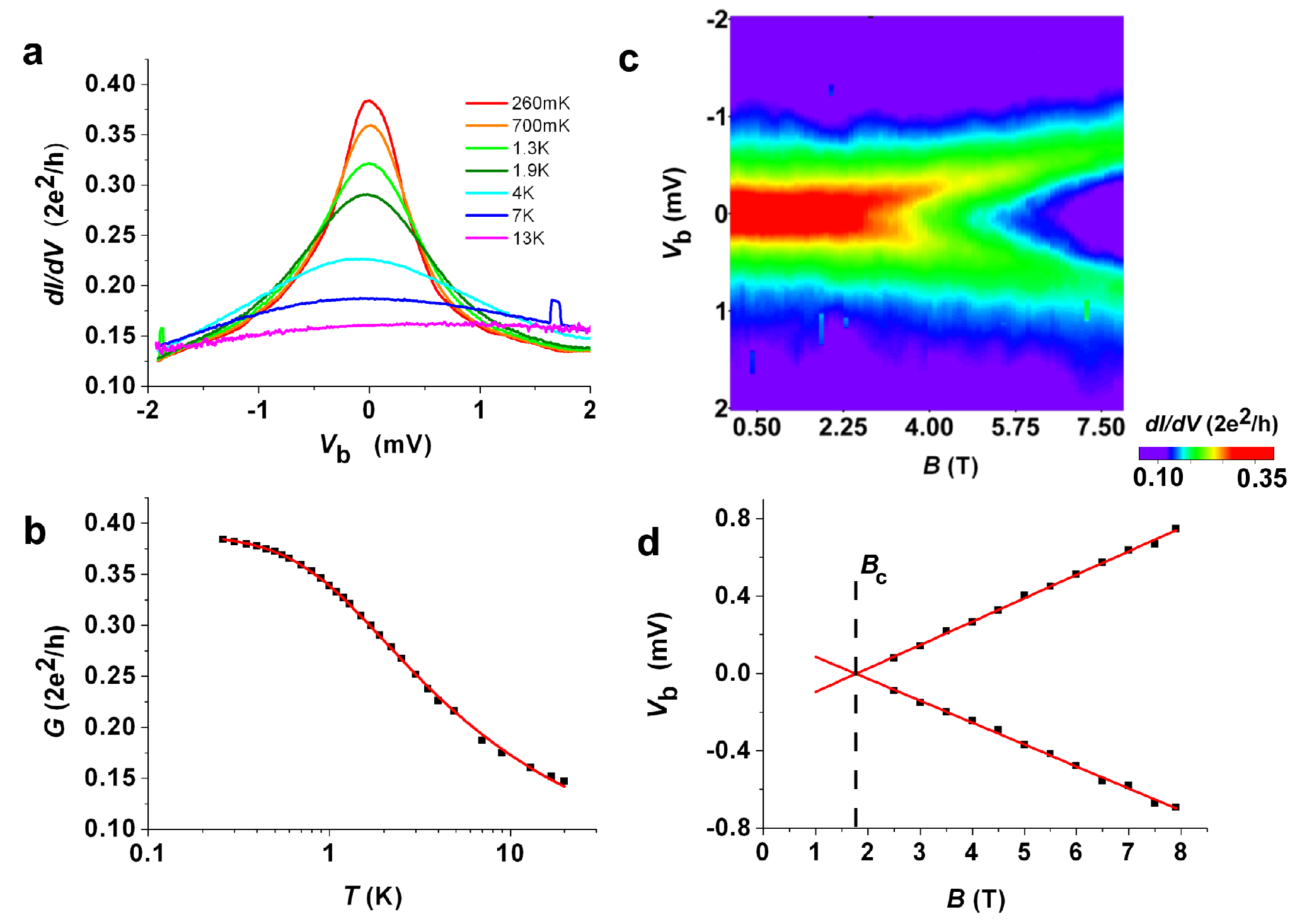}
\caption{ {\bf Temperature and magnetic study of fully-screened spin 
$S=1/2$ Kondo effect.} {\bf a,} Evolution of differential conductance versus bias
voltage for several temperatures from 260~mK to 20~K. {\bf b,} Temperature
dependence of the conductance at $V_{\rm b}=0$~mV extracted from the
{\bf a} panel, with a fit to the empirical formula proposed by
Goldhaber-Gordon {\it{et al.}}~\cite{Goldhaber1998bis} which gives $T_{\rm
K}=4.46$~K. {\bf c,} Differential conductance ($\partial I/\partial V$) map
versus bias voltage ($V_{\rm b}$) and magnetic field $B$. {\bf d,}
Position of the Kondo peaks extracted from the {\bf c,} plot. The linear
extrapolation of the peak position gives a critical field $B_{\rm
c}=1.78$~T.}
\label{figspin1_2}
\end{figure}

We demonstrate in the following that all these features are
very cleanly observed in our $\rm{C_{60}}$ quantum dot, thereby giving
the basis
of the study presented in the main paper of the singlet-triplet
transition for the even charge valley (measured with the {\it same}
device). In Fig.~S.~\ref{figspin1_2}a we plot the differential
conductance versus bias voltage. When the temperature is lowered, the
height of the peak increases due to the Kondo effect. We estimate the value of $T_{\rm K}$ by
measuring the half width at half maximum~(HWHM) of the peak for $T
\ll T_{\rm K}$. At $T=260$~mK, we find
$V_{\mathrm{b}}^\mathrm{HWHM} = 380$~$\rm{\mu V}$, corresponding to $T_{\rm
K}=4.42$~K.
A second and more precise way to find $T_{\rm K}$ is to fit the
temperature evolution of the conductance at zero bias. The precise
shape of this curve is universal (up to the value of energy scale
$T_{\rm K}$), and can be calculated by Numerical Renormalization
Group (NRG) theory~\cite{Costi2000}. An empirical formula based on this
calculation was found by Goldhaber-Gordon {\it{et al.}}~\cite{Goldhaber1998bis} and
is used to fit the experimental data, as shown in 
Fig.~S.~\ref{figspin1_2}b, where we find $T_{\rm K}$~=~4.46~K.
An additional method is to use the
magnetic field dependence of the conductance~\cite{Kogan2004}.
The Zeeman effect competes with the Kondo resonance so that
a non-equilibrium Kondo peak appears roughly at $V_{\rm{b}}=g\mu
_{\rm{B}}B$, as shown in Fig.~S.~\ref{figspin1_2}c). This splitting is
predicted to appear for 
$g\mu_{\rm{B}}B_{\rm{c}}=0.5{k_{\rm{B}}}{T_{\rm{K}}}$~\cite{Costi2000}.
In Fig.~S.~\ref{figspin1_2}d
we linearly interpolate the position of these peaks and find
$B_{\rm{c}}$~=~1.78~T, which yields $T_{\rm{K}}$~=~4.78~K. The value 
of
$T_{\rm{K}}$ obtained with the three different methods are consistent 
and demonstrate a well defined $T_{\rm{K}}$.

\section{Non-equilibrium singlet-triplet Kondo effet on the singlet side}

In this section, we demonstrate, in our $\rm{C_{60}}$ molecular 
junction, an
effect recently reported by Paaske {\it{et al.}} in a carbon nanotube quantum
dot~\cite{Paaske2006}, namely the non-equilibrium singlet-triplet Kondo effect.
These authors were the first to clearly identify sharp finite voltage bias
features as a Kondo effect and not as simple cotunneling via excited states.
The main idea behind Kondo physics is the existence of a degeneracy, which is
lifted by the conduction electrons. This is clearly the case for a quantum dot
with only one electron on the last orbital, leading to a doubly degenerate
spin $S=1/2$.
For a quantum dot with two electrons and two nearly degenerate orbital
levels, two different kinds of magnetic states occur: a singlet and a
triplet. Depending on $\delta E$ the energy difference between the
two orbital levels and $J$ the strength of the ferromagnetic
coupling between the two electrons, the splitting between the triplet and
the singlet can in principle be tuned, and eventually brought to zero,
leading to the so-called singlet-triplet Kondo effect~\cite{Schmid2000}.
However the singlet is in most situations the ground state, leaving the triplet
in an excited state, thus suppressing the Kondo effect. Kondo signatures can
nevertheless be observed by tuning the degeneracy in a magnetic
field~\cite{Sasaki2000, Nygard2000, Liang2002}.
Another way to retrieve the degeneracy is to apply a bias voltage,
although it is of course more delicate to preserve the quantum coherence
necessary to Kondo correlations. Indeed, finite-bias features
clearly linked to magnetic excitations were observed in
2DEGs~\cite{Zumbuhl2004}, carbon nanotubes~\cite{Nygard2000, Liang2002,
Babic2004,Quay2007} and even recently in an OPV5 molecule~\cite{Osorio2007}. 
However, only the study reported by Paaske {\it{et al.}}~\cite{Paaske2006}
was able to identify a clear non-equilibrium Kondo effect.
Their first observation was the occurence of sharp peaks in the differential
conductance for both positive and negative bias voltage, very different from
the cusps usually associated to cotunneling.
Secondly the height of these peaks decreased logarithmically with temperature,
which is another typical signature of Kondo correlations.
Finally the shape of the peaks could be well accounted for in a non-equilibrium
Kondo calculation, while a simple cotunneling model failed to reproduce the
data.

\def\figurename{Fig. S.}
\begin{figure}
 \includegraphics[width=16cm]{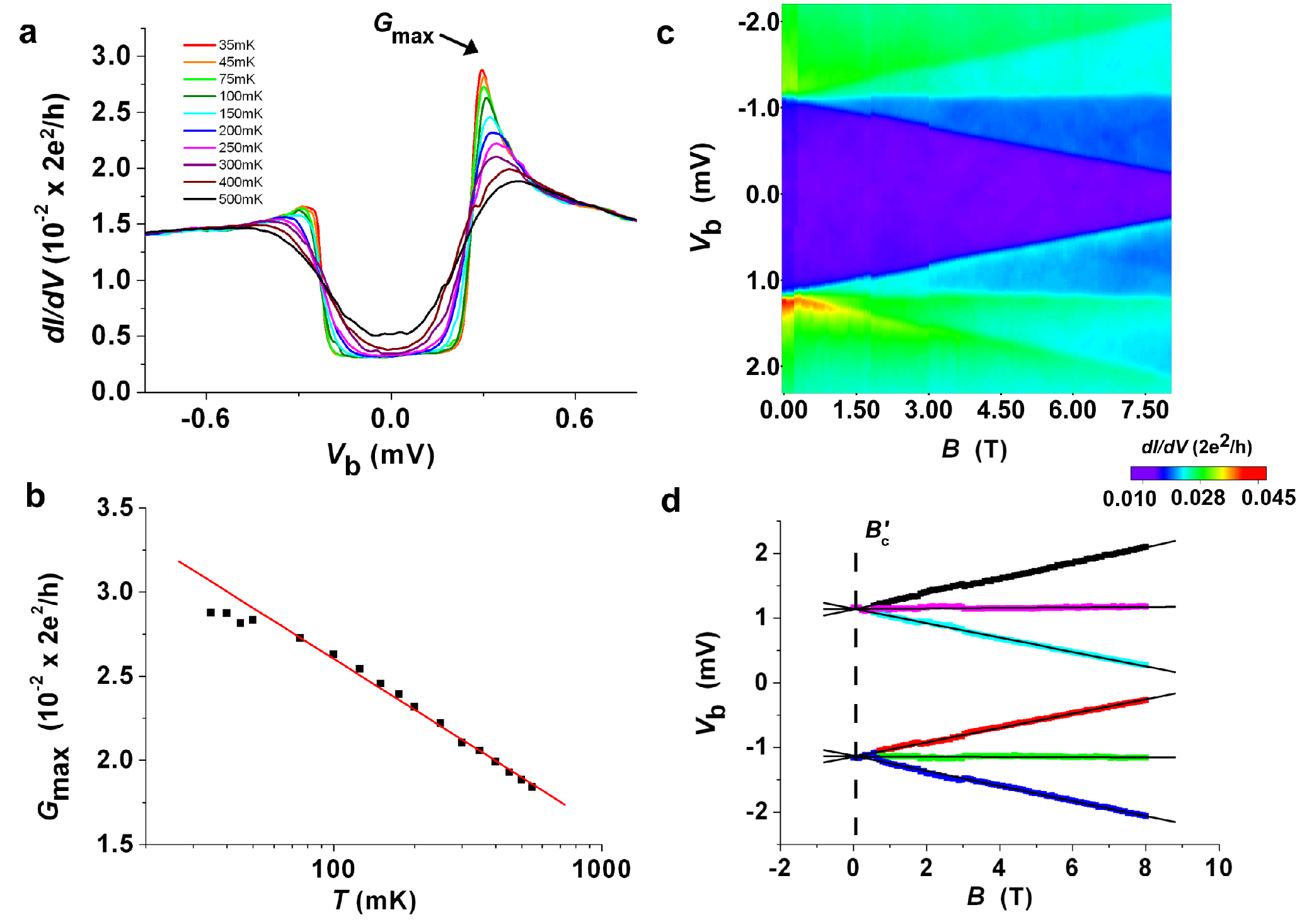}
 \caption{ {\bf Non-equilibrium singlet-triplet Kondo effect on the 
 singlet side.} {\bf a,} Differential conductance versus bias voltage for temperature from
35~mK~(black) to 500~mK~(pink) at fixed $V_{\rm{g}}=1.79$~V. {\bf b,} Evolution of
the "positive $V_{\rm{b}}$" peak height in a) with temperature on a
logarithmic scale, which can be linearly-fitted on nearly a decade.
{\bf c,} Differential conductance map as a function of bias voltage and
magnetic field at fixed $V_{\rm{g}}=1.64$~V. {\bf d,} Position of the excited
triplet-peaks extracted from {\bf c}. The linear fits demonstrate that
the non-equilibrium singlet-triplet Kondo peaks split at a finite
magnetic field $B_{\rm{c}}^{\prime}=50$~mT.}
\label{figtriplethorsequilibre}
 \end{figure}

These striking features are also present in our experiment, for the
case of an even charge state into the $\rm{C_{60}}$ molecule. We
focus here on the singlet side, but similar results are observed for
the Kondo satellites on the triplet side (see main text). Indeed,
while the conductance at low bias is suppressed when the spin state
of the system is a singlet, a clear finite-bias peak grows by
decreasing temperature as shown in
Fig.~S.~\ref{figtriplethorsequilibre}a. In addition, the amplitude of
the positive bias peak decreases logarithmically about a decade
(Fig.~S.~\ref{figtriplethorsequilibre}b), showing a clear signature of
the non-equilibrium singlet-triplet Kondo effect. The magnetic field
dependence of the differential conductance presented in
Fig.~S.~\ref{figtriplethorsequilibre}c is also very interesting. This
plot, which was not numerically treated, shows the Zeeman splitting
between the three triplet states at both positive and negative bias.
The positions of those peaks are reported on
Fig.~S.~\ref{figtriplethorsequilibre}d and a linear fit is applied to
each line, with a very good accuracy which enables us to determine, 
firstly, a
critical field $B_{\rm{c}}^{\prime}$ of 50~mT before the splitting occurs, and secondly, a Lande
factor $g=2\pm 0,1$. The existence of a critical field for the
splitting of the {\it zero-bias} anomaly is well-documented in the
case of the Kondo effect in equilibrium (see also our study for the
spin $1/2$ of section~\ref{sec:screened}). To our knowledge, these
data are the first observation of this effect for the {\it finite
bias} satellites associated to the non-equilibrium singlet-triplet
Kondo effect. By applying the relation found by Costi~\cite{Costi2000} for
the spin $S=1/2$ case, we estimate the
Kondo temperature $T_{\rm{K}}=130$~mK. This value must be taken as
an approximation since the spin $S=1/2$ model does certainly
not apply quantitatively here, and also the base temperature $T=35$~mK was
not much smaller than $T_{\rm{K}}$. Again, because the
charging energy of a $\rm{C_{60}}$ molecule is twenty times larger
than that of a carbon nanotube quantum dot, we are able to access
relatively high Kondo scale (the out-of-equilibrium Kondo
temperature was estimated to be 2~mK by Paaske {\it{et al.}} in their
device~\cite{Paaske2006}).

\section{Singlet-triplet transition: low versus very low temperature}

\begin{figure}
 \includegraphics[width=16cm]{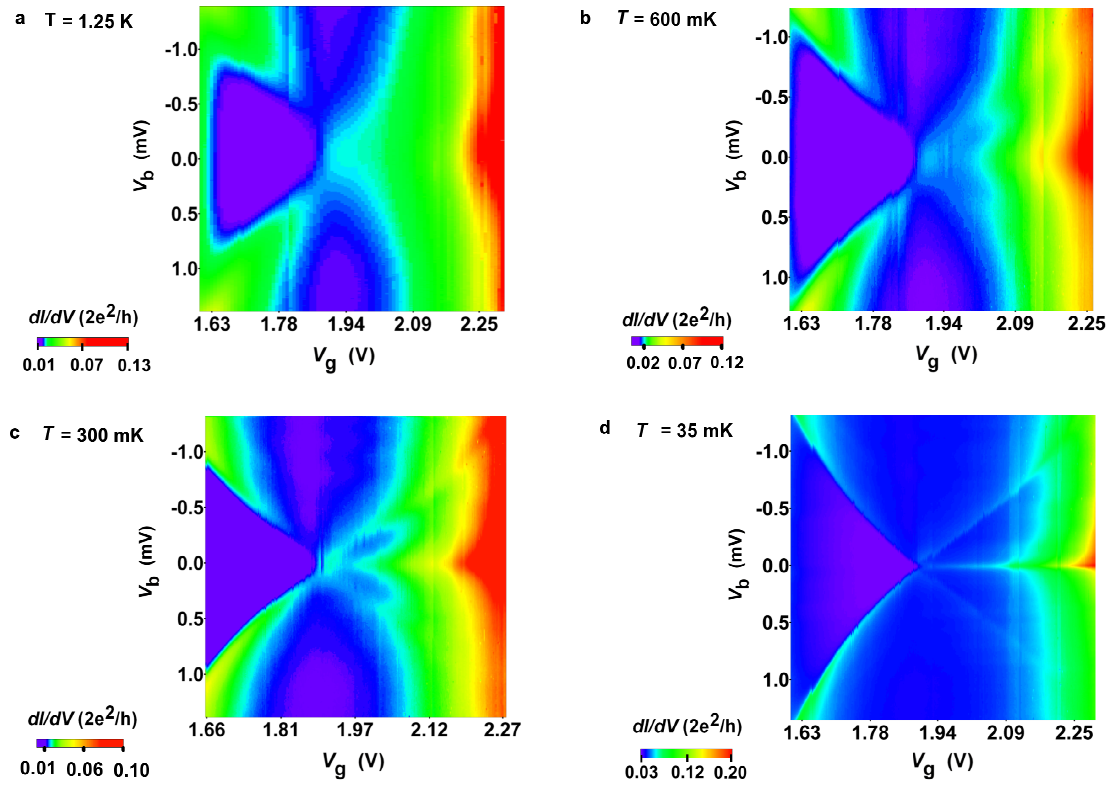}
 \caption{{\bf Differential conductance maps versus bias voltage and 
 gate voltage.}
 The maps at $T=35$~mK~({\bf d}), $T=300$~mK~({\bf c}) and 
 $T=600$~mK~({\bf b}) were measured with
 the same parameters ($V_{\rm{AC}}=10$~${\mu V}$) whereas the one at
 $T=1.25$~K~({\bf a}) was measured with an higher AC excitation
 amplitude ($V_{\rm{AC}}=30$~$\rm{\mu V}$). }
\label{figLMTemp}
 \end{figure}

In this section, we compare our $\rm{C_{60}}$ quantum dot results on the singlet-triplet
transition to previous studies on different quantum dot systems (2DEG or carbon
nanotubes), and argue that the temperature required to observe a critical
Kondo behavior was certainly too low in these previous experiments.
The first important fact is that single molecules offer typically higher energy
scales (charging energy, Kondo temperature...) due to their extremely small size.
It is for example possible to study very well the spin $S=1/2$ Kondo effect at
liquid helium temperature in a $\rm{C_{60}}$ molecular junction~\cite{Yu2004, Parks2007}.
The second crucial ingredient in our experiment was the development of the
electromigration technique in a highly filtered dilution fridge, as discussed
in section~\ref{sec:setup}, which allowed us to reach temperatures well below the
"high-energy" Kondo scale. Both points are well illustrated in Fig.~S.~\ref{figLMTemp},
which shows the conductance maps at four different temperatures. 
Fig.~S.~\ref{figLMTemp}a
corresponds to measurements at $T=1.25$~K. At this temperature, we 
observe a dip at the singlet side, but
a single broad peak at the triplet side. This is reminiscent of the data
reported by Kogan {\it{et al.}}~\cite{Kogan2003} or Quay {\it{et al.}}~\cite{Quay2007},
which showed a featureless change of behavior near the singlet-triplet
crossing. On the contrary, our datas at the much lower base
temperature $T=35$~mK
demonstrate more complex 
features~(Fig.~S.~\ref{figLMTemp}d), that we associated to a singlet-triplet
quantum phase transition (see main text). In our point of view, the small Kondo
temperatures obtained in other quantum dot systems prevented to observe those
effects, even using a dilution refrigerator.

\def\figurename{Fig. S.}

\section{Temperature dependence of the zero-bias conductance}

\begin{figure}
\includegraphics[width=16cm]{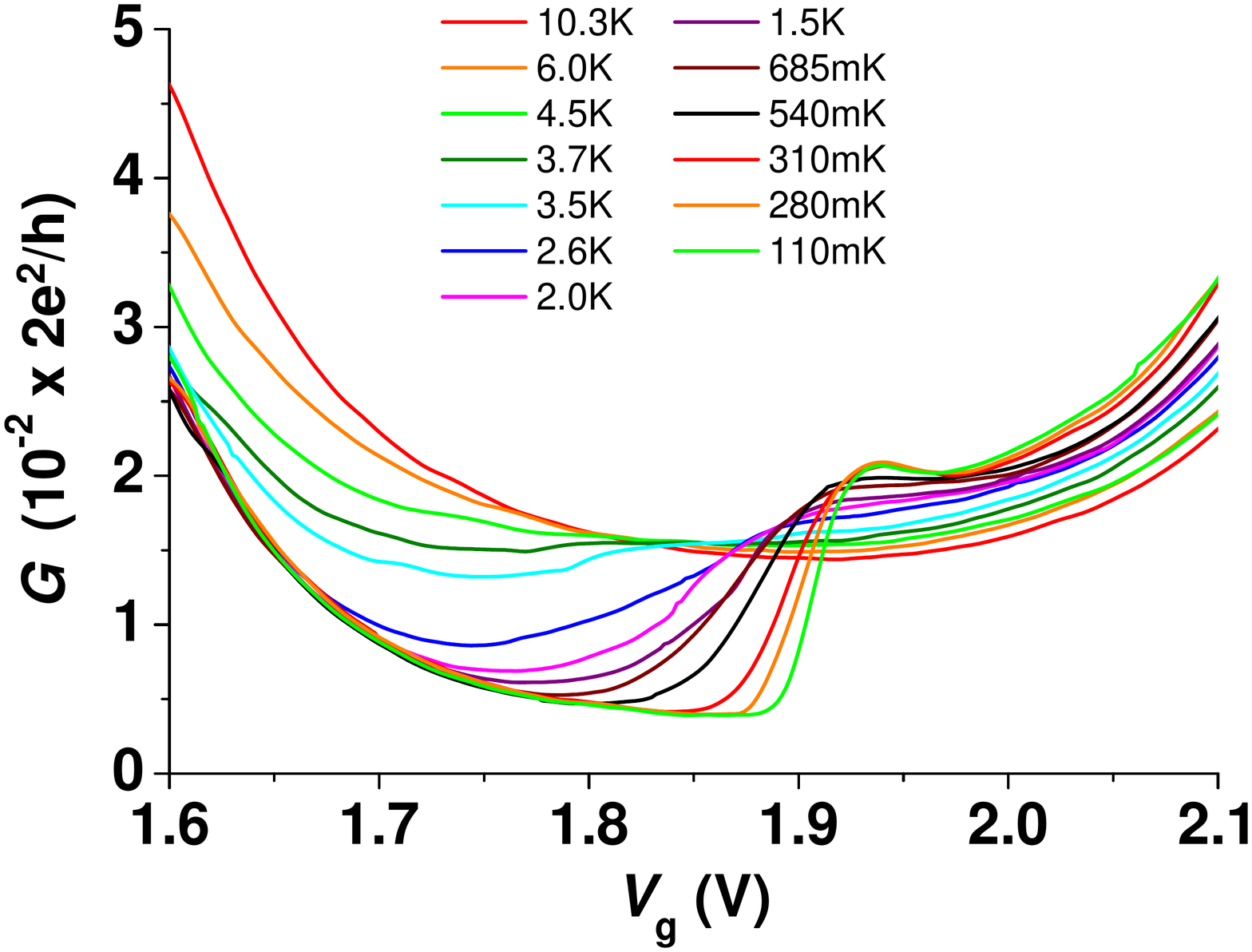}
\caption{{\bf Conductance at zero-bias versus gate voltage for
 temperature from 110~mK to 10.3~K.}}
 \label{figConductanceTemp}
 \end{figure}

The singlet-triplet transition in quantum dots was widely
studied~\cite{Nygard2000, Schmid2000, Sasaki2000, Babic2004, Quay2007,
Kogan2003}. In those cases, a clear maximum of conductance appeared
when the singlet and the triplet states were driven through
degeneracy by magnetic field or gate voltage. On Fig.~S.~\ref{figConductanceTemp}, the
zero bias conductance is presented for different temperatures as a function
of gate voltage. At $T=10.3$~K we cannot discriminate the
singlet from the triplet. Lowering the temperature, conductance
decreases at the singlet side whereas it increases at the triplet
side, no peak appearing at the singlet to triplet transition.

\def\figurename{Fig. S.}

One remarkable aspect of
this measurement is the absence of an enhancement of the zero bias 
conductance at the singlet to triplet transition.
The lack of such gate-induced Kondo effect points towards a predominent coupling
of the $\rm{C}_{60}$ QD to a single screening channel, leading to 
 a strong proof of the observation of a quantum phase
transition.

\section{Statistics and reproducibility of the results}

\def\figurename{Fig. S.}
\begin{figure}
\includegraphics[width=16cm]{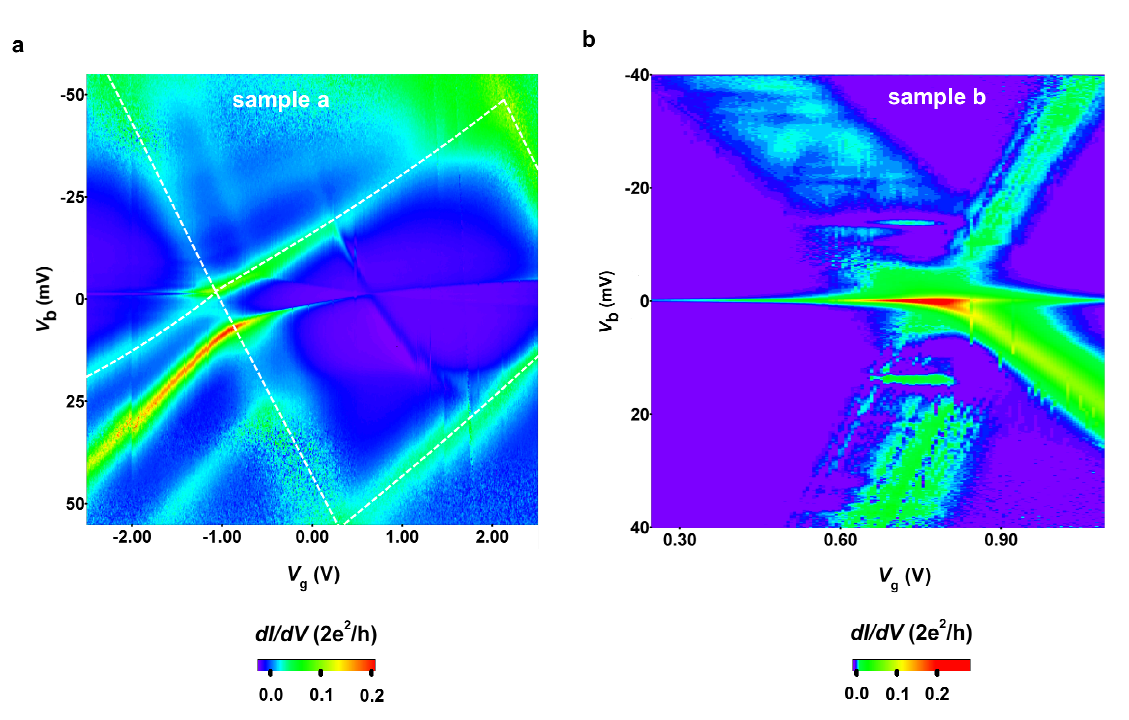}
\caption{{\bf Colour-scale map of the differential conductance 
$\partial I/\partial V$ as a function of bias voltage Vb and gate 
voltage Vg at $40$~mK and zero magnetic field.} {\bf a,} This 
single-molecule quantum dot exhibits an excitation of the order of 
30~meV and a fine gate-tuning of the singlet to triplet energy 
difference inside the Coulomb 
diamond (dooted rectangle). {\bf b,} This device clearly exhibits a 
large charging energy and a spin S~=~1/2 and S~=~1 Kondo 
effects.}
 \label{stat}
 \end{figure}

Preparation of the single-molecule quantum dot studied in the 
article was realized by blow drying 
a dilute toluene solution of the $\rm{C_{60}}$ molecule onto a gold 
nano-wire 
realized on an $\rm{Al/Al_{2}O_{3}}$ back gate. Before blow 
drying the solution, the electrodes were cleaned with acetone, ethanol, 
isopropanol solution and oxygen plasma. As it is known that even if the 
electromigration procedure is 
well controlled, there is always a possibility of realizing a gold 
agregat~\cite{Houck2005} containing few atoms, we studied several 
jonctions prepared 
within the same procedure with a toluene solution only. 
In our opinion, it is relevant to state here that an "interesting" 
device to investigate must show at least one order of 
magnitude change in the current caracteristics as a function of 
the gate 
voltage for a 1~mV voltage bias, and a charging 
energy greater than 20~meV. Within these drastic restrictions, we tested 
38 bared junction 
with a toluene solution and 51 with a dilute $\rm{C_{60}}$ toluene solution. 
If 3 bared junction 
showed one order of magnitude changes in the current as a function of the 
gate voltage after electromigration, 
only 2 had a charging energy higher than 20~meV, and only 1 of 
those 2 exhibited a zero bias anomaly. These transport structures were also 
not very well defined.
For junctions prepared with a diluted $\rm{C_{60}}$ toluene solution, we 
measured 7 
jonctions out of 51 with one order of magnitude changes in the 
current as a function of 
gate voltage, and 6 of those 7 had a charging energy 
higher than 20~meV and exhibited zero bias anomalies. 

We present in this section measurements on two different samples, that we 
performed in the same conditions as the device presented in the 
article. The first one 
presented in Fig.\ref{stat}.b exhibits a large 
charging energy, spin S~=~1/2 and S~=~1 Kondo effects, and 
multiple excited states. We 
did not investigate further this measurement because we could 
not discriminate, unlike the device presented in the main paper, the 
different triplet states by applying a magnetic field. However, on the 
right side of the degeneracy point, the zero bias anomaly splitted at 
a magnetic field of the order of 1.9~T, while the zero bias anomaly on 
the left side splitted for a magnetic field of the order of 100~mT. 
The second measurement we present in Fig.\ref{stat}.a exhibits the same 
Kondo behavior than the 
single-molecule quantum dot presented in the article, and an excitation 
that may be related 
to the vibrational excitation energy of a $\rm{C_{60}}$ single molecule 
connected to gold electrodes. We also clearly observe the singlet to triplet 
out of equilibrium Kondo effect, and a gate-tuning of the energy 
difference between the singlet and the triplet, similar to the 
single-molecule quantum dot of the article. However, we do not measure 
an underscreened Kondo effect. We assume that the Kondo temperature 
is too low, because of a weaker coupling to the electrodes, to measure this 
effect, or, in contrast with the 
device exhibiting the quantum phase transition, that we are in the 
2-screening channel limit. This device, currently under investigation, 
exhibits other kind of 
excitations in the Coulomb blockade diamond, which are not, so far, 
well understood. But it is important to state here that some of the 
physics that enabled us to observe the quantum phase transition in the 
single-molecule quantum dot presented in the paper is reproducible, 
and that we never measured such behaviors in junctions prepared 
without a dilute $\rm{C_{60}}$ toluene solution.


\end{document}